\documentclass[12pt]{article}

\usepackage{cite}
\usepackage{amsmath}
\usepackage{amssymb}
\usepackage{pstricks}

\newcommand{\mycaption}[1]{\caption{\sl #1}}

\makeatletter
\def\section{\@startsection {section}{1}{\z@}{+3.0ex plus +1ex minus
  +.2ex}{2.3ex plus .2ex}{\large\bf\boldmath}}
\def\subsection{\@startsection{subsection}{2}{\z@}{+2.5ex plus +1ex
minus +.2ex}{1.5ex plus .2ex}{\normalsize\bf\boldmath}}
\def\subsubsection{\@startsection{subsubsection}{3}{\z@}{+3.25ex plus
 +1ex minus +.2ex}{1.5ex plus .2ex}{\normalsize\it}}

\begin{document}
\thispagestyle{empty}

\def\thefootnote{\fnsymbol{footnote}}

\begin{flushright}
\end{flushright}

\vspace{1cm}

\begin{center}

{\large {\bf TVID: Three-loop Vacuum Integrals from Dispersion relations}}
\\[3.5em]
{\large
Stefan Bauberger$^1$, Ayres~Freitas$^2$
}

\vspace*{1cm}

{\sl
$^1$ Hochschule f\"ur Philosophie,
Philosophische Fakult\"at S.J.,
Kaulbachstr.\ 31,\\ 80539 M\"unchen, Germany \\[1ex]
$^2$ Pittsburgh Particle-physics Astro-physics \& Cosmology Center
(PITT-PACC),\\ Department of Physics \& Astronomy, University of Pittsburgh,
Pittsburgh, PA 15260, USA
}

\end{center}

\vspace*{2.5cm}

\begin{abstract}

TVID is a program for the numerical evaluation of general three-loop vacuum
integrals with arbitrary masses. It consists of two parts. An algebraic module,
implemented in Mathematica, performs the separation of the divergent pieces of
the master integrals and identifies special cases. The numerical module,
implemented in C, carries out the numerical integration of the finite pieces. In
this note, the structure of the program is explained and a few usage examples
are given.

\end{abstract}

\setcounter{page}{0}
\setcounter{footnote}{0}

\newpage


\section{Overview}

\paragraph{Program name and version:} TVID, version 1.0 (December 2016).

\paragraph{System requirements:} {\sc Linux}-compatible platform; GNU C compiler {\sc gcc
4.4} or similar; {\sc Mathematica 10.x}.

\paragraph{Copyright:} The TVID source code may be freely used and incorporated
into other projects, but the authors ask that always a reference to this
document and to Ref.~\cite{paper} be included.

\paragraph{External code elements included:} TVID uses the Gauss-Kronrod routine
QAG from the {\sc Quadpack} library \cite{quadpack}, translated into C++, and
the C++ package {\sc doubledouble} for 30 digit floating point
arithmetic \cite{quad}.

\paragraph{Code availability:} The TVID source code is available for download at
\newline {\tt http://www.pitt.edu/\~{}afreitas/}.



\section{Introduction}

\noindent
TVID is a public computer code for the efficient and robust numerical evaluation
of a generic basis of master integrals for the three-loop vacuum diagrams. It is
based on an approach using dispersion relations, which is described in
Ref.~\cite{paper}. The method makes use of ideas which were previously
developed for the numerical evaluation of two-loop self-energy integrals
\cite{disp2}.

Analytical results for three-loop vacuum integrals are known for cases with one
or two independent mass scales
\cite{ana,Broadhurst:1998rz,Chetyrkin:1999qi,Grigo:2012ji}. However, for fully
general mass configurations, only numerical approaches are currently available.
Besides the technique used in TVID, an alternative numerical method based on
differential equations has been discussed in Ref.~\cite{3vac}.

TVID provides results for the divergent and finite pieces of the three-loop
vacuum master integrals. For some applications it may be necessary to evaluate the master
integrals to higher orders in $\epsilon$, which requires a non-trivial extension
of the method of Ref.~\cite{paper}.

\begin{figure}[b]
\begin{center}
\psset{linewidth=1pt}
\psset{dotsize=5pt}
\begin{tabular}{p{3cm}p{3cm}p{3cm}}
\\[1em]
&& \\[-.2cm]
\begin{center}
\pscircle(0,0){1}%
\psdot(0,1)%
\psdot(0,-1)%
\psdot(-1,0)%
\psarc(-1,0){1.414}{-45}{45}%
\psarc(1,0){1.414}{135}{-135}%
\rput[r](-0.9,0.7){\small 1}%
\rput[r](-0.9,-0.7){\small 1}%
\rput[r](-0.5,0){\small 2}%
\rput[r](0.3,0){\small 3}%
\rput[l](1.05,0){\small 4}%
\end{center}
&
\begin{center}
\pscircle(0,0){1}%
\psdot(-0.707,0.707)%
\psdot(0,-1)%
\psdot(0.707,0.707)%
\psline(0,-1)(-0.707,0.707)%
\psline(0,-1)(0.707,0.707)%
\rput[r](-1,-0.4){\small 1}%
\rput[r](-0.45,-0.2){\small 2}%
\rput[l](1,-0.4){\small 4}%
\rput[l](0.45,-0.2){\small 3}%
\rput[t](0,0.9){\small 5}%
\end{center}
&
\begin{center}
\pscircle(0,0){1}%
\psdot(-0.866,0.5)%
\psdot(0,-1)%
\psdot(0.866,0.5)%
\psdot(0,0)%
\psline(0,0)(-0.866,0.5)%
\psline(0,0)(0.866,0.5)%
\psline(0,0)(0,-1)%
\rput[r](-1,-0.4){\small 1}%
\rput[r](-0.4,0){\small 2}%
\rput[l](1,-0.4){\small 4}%
\rput[l](0.4,0){\small 3}%
\rput[t](0,0.9){\small 5}%
\rput[l](0.1,-0.5){\small 6}%
\end{center}
\\[0.6cm]
\centering $U_4$ & \centering $U_5$ & \centering $U_6$
\end{tabular}
\end{center}
\vspace{-2.5ex}
\mycaption{Basic master integral topologies used by TVID. The dot
indicates a propagator that is raised to the power 2.
\label{fig:diag1}}
\end{figure}

\medskip\noindent
The set of master integrals is shown in Fig.~\ref{fig:diag1} and defined by
\begin{align}
&U_4(m_1^2,m_2^2,m_3^2,m_4^2) \equiv M(2,1,1,1,0,0), \\
&U_5(m_1^2,m_2^2,m_3^2,m_4^2,m_5^2)  \equiv M(1,1,1,1,1,0), \\
&U_6(m_1^2,m_2^2,m_3^2,m_4^2,m_5^2,m_6^2) \equiv M(1,1,1,1,1,1), 
\end{align}
where
\begin{align}
M(\nu_1,\nu_2,\nu_3,\nu_4,\nu_5,\nu_6;\, &m_1^2,m_2^2,m_3^2,m_4^2,m_5^2,m_6^2)
\nonumber \\
= i\frac{e^{3\gamma_{\rm E}\epsilon}}{\pi^{3D/2}}
  & \int d^Dq_1\, d^Dq_2\, d^Dq_3 \;
  \frac{1}{[q_1^2-m_1^2]^{\nu_1} [(q_1-q_2)^2-m_2^2]^{\nu_2}} \nonumber \\
&\;\times
  \frac{1}{[(q_2-q_3)^2-m_3^2]^{\nu_3} [q_3^2-m_4^2]^{\nu_4}
	   [q_2^2-m_5^2]^{\nu_5} [(q_1-q_3)^2-m_6^2]^{\nu_6}}\,, \label{eq:m}
\end{align}
and $\epsilon = (4-D)/2$, $D$ is the number of dimensions in dimensional
regularization. For generic values of the mass parameters,
any three-loop vacuum integral can be reduced to a linear combination of $U_4$,
$U_5$ and $U_6$ functions with the help of integration-by-parts
identities~\cite{ibp}. 

The master integrals $U_4$, $U_5$ and $U_6$ have the following symmetry
properties:
\begin{itemize}
\item
$U_4(m_1^2,m_2^2,m_3^2,m_4^2)$ is invariant under arbitrary permutations of
$m_{2,3,4}$.
\item
$U_5(m_1^2,m_2^2,m_3^2,m_4^2,m_5^2)$ is invariant under the replacements $\{m_1
\leftrightarrow m_2\}$, $\{m_3 \leftrightarrow m_4\}$, and 
$\{m_1 \leftrightarrow m_3, \, m_2 \leftrightarrow
m_4\}$, as well as any combination thereof.
\item
$U_6(m_1^2,m_2^2,m_3^2,m_4^2,m_5^2,m_6^2)$ is invariant under the replacements
$\{m_2 \leftrightarrow m_3,\, m_1 \leftrightarrow m_4\}$,
$\{m_2 \leftrightarrow m_6,\, m_4 \leftrightarrow m_5\}$,
$\{m_1 \leftrightarrow m_6,\, m_3 \leftrightarrow m_5\}$,
and any combination thereof.
\end{itemize}


\section{Structure and usage of the program}
\label{sc:prog}

\subsection{Numerical part}
\label{sc:num}

\noindent
The numerical part of TVID is programmed in C and evaluates the finite remainder
functions defined in Ref.~\cite{paper}. These are
\begin{align}
U_{4,\rm sub}(m_1^2,m_2^2,m_3^2,m_4^2) 
&= U_4(m_1^2,m_2^2,m_3^2,m_4^2) - U_4(m_1^2,0,m_3^2,0) - U_4(m_1^2,0,0,m_4^2) \nonumber \\ 
&\quad+ 2\, U_4(m_1^2,0,0,0) \label{eq:u4sub}
\end{align}
for $m_1 \neq 0$, see section 3.1 in Ref.~\cite{paper};
\begin{align}
&U_{4,\rm sub,0}(m_2^2,m_3^2,m_4^2) \nonumber \\
&\quad= U_4(0,m_2^2,m_3^2,m_4^2) - \Bigl [B_0(0,0,0)-B_0(0,\delta,\delta)-\log
\tfrac{\delta^2}{m_2^2} \Bigr ] T_3(m_2^2,m_3^2,m_4^2) \nonumber \\[-1ex] 
&\quad\quad-\log \tfrac{\delta^2}{m_2^2} \sum_{i=2}^4 T_3(m_i^2,0,0) \\
&\quad= U_{4,\rm sub}(\delta^2,m_2^2,m_3^2,m_4^2) + \log \tfrac{\delta^2}{m_2^2} \Bigl [
T_3(m_2^2,m_3^2,m_4^2) - \sum_{i=2}^4 T_3(m_i^2,0,0) \Bigr ],
\end{align}
where $\delta>0$ is an infinitesimally small parameter, see section 3.2 in Ref.~\cite{paper};
\begin{align}
U_{5,\rm sub}(m_1^2,m_2^2,m_3^2,m_4^2,m_5^2) &= M(2,1,1,2,1,0;\,m_1^2,m_2^2,m_3^2,m_4^2,m_5^2)
\end{align}
for the finite part of the generic $U_5$ case, where $M$ is defined in
eq.~\eqref{eq:m}, see section 4.1 in Ref.~\cite{paper};
\begin{align}
&U_{5,\rm sub,0}(m_3^2,m_4^2,m_5^2) = M(1',1,1,2,1,0;\,m_4^2,0,m_3^2,m_4^2,m_5^2) \nonumber \\
&\qquad \begin{aligned}[b]
=  i\frac{e^{3\gamma_{\rm E}\epsilon}}{\pi^{3D/2}}
  \int d^Dq_1\, d^Dq_2\, d^Dq_3 \;
  &\frac{1}{q_1^2[q_1^2-m_4^2] (q_1-q_2)^2} \\
 &\times \frac{1}{[(q_2-q_3)^2-m_3^2] [q_3^2-m_4^2]^2
	   [q_2^2-m_5^2]}\,,
\end{aligned}
\end{align}
for the special case $U_5(0,0,m_3^2,m_4^2,m_5^2)$, see section 4.3 in
Ref.~\cite{paper}; and
\begin{align}
U_{6,\rm sub}(m_1^2,m_2^2,m_3^2,m_4^2,m_5^2,m_6^2) &=
U_6(m_1^2,m_2^2,m_3^2,m_4^2,m_5^2,m_6^2) \nonumber \\ &\quad-
U_6(m_6^2,m_6^2,m_6^2,m_6^2,m_6^2,m_6^2),
\end{align}
where, without loss of generality, it has been assumed that $m_6 \geq m_i$
for $i=1,\dots,5$, see section 5 in
Ref.~\cite{paper}.

\medskip\noindent
The numerical code is called with the command
\begin{quote}
{\tt ucall} {\it infile} {\it outfile}
\end{quote}
where {\it infile} is the name of the input file, and {\it outfile} is the name
of the file where the results shall be placed. {\it infile} may contain a list
of lines, separated by line breaks, where each line has the form
\begin{quote}
{\it fname} \ {\it massA} \ {\it massB} \ \dots
\end{quote}
Here {\it fname} is the name of the function to be evaluated, see
Tab.~\ref{tab:fnames}, and {\it massA}, {\it massB}, etc.\ are the numerical
mass parameters supplied. For example,
\begin{quote}
\tt U4 \ 1 \ 2 \ 3 \ 4 \\
\tt U50 \ 1.5 \ 2.5 \ 0.5
\end{quote}
asks for the evaluation of $U_{4,\rm sub}(1,2,3,4)$ and of
$U_{5,\rm sub,0}(1.5, 2.5,0.5)$. When $\tt ucall$ is completed, it fills {\it outfile} with
a list of the numerical results, again separated by line breaks. For instance, the
example above will return
\begin{quote}
\tt 
-5.555128856244808e1 \\
-1.537493378796251
\tt 
\end{quote}
Internally, the numerical code uses the Gauss-Kronrod routine
QAG from the {\sc Quadpack} library \cite{quadpack} to evaluate the dispersion
integrals. This routine has been translated into C++ from the original FORTRAN
code, and amended to facilitate 30 digit floating point arithmetic from the
package {\sc doubledouble} \cite{quad}.
\begin{table}[tb]
\begin{center}\
\begin{tabular}{|l|l|l|l|}
\hline
& & Symbol {\it fname} & Symbol used \\
Function & Mass parameters & used by numerical & by algebraic \\
& & code {\tt ucall} & Mathematica code \\
\hline
$U_{4,\rm sub}$ & $m_1,m_2,m_3,m_4$ & \tt \ U4 & \tt \ U4sub \\
$U_{4,\rm sub,0}$ & $m_2,m_3,m_4$ & \tt \ U40 & \tt \ U4sub0 \\
$U_{5,\rm sub}$ & $m_1,m_2,m_3,m_4,m_5$ & \tt \ U5 & \tt \ M21121 \\
$U_{5,\rm sub,0}$ & $m_3,m_4,m_5$ & \tt \ U50 & \tt \ M1p1121 \\
$U_{6,\rm sub}$ & $m_1,m_2,m_3,m_4,m_5,m_6$ & \tt \ U6 & \tt \ U6sub \\
\hline
\end{tabular}
\end{center}
\vspace{-2.5ex}
\mycaption{Symbols for basic finite remainder functions used in the numerical
and algebraic parts of TVID.
\label{tab:fnames}}
\end{table}


\subsection{Algebraic part}
\label{sc:al}

The algebraic part of TVID runs in Mathematica 10 \cite{mathematica}. It
performs the separation of divergent and finite pieces of the master integrals
and identifies cases that a require special treatment. The program is loaded
into a Mathematica session with
\begin{quote}
\tt << i3.m
\end{quote}
It contains two basic functions, {\tt PrepInt} and {\tt UCall}.

\medskip\noindent 
{\tt PrepInt} takes any $U_4$, $U_5$ or $U_6$ function as input and splits them
into divergent terms and the finite remainder functions introduced in the
previous subsection. It returns a series expansion in $\epsilon$. For example
\begin{quote}
\tt\small \rule{0mm}{0mm}\\[-2ex]
In[2]:=~PrepInt[U4[1.,2.,2.,1.]]~\\
\\
\rule{0mm}{0mm}~~~~~~~~1.66667~~~2.61371~~~7.48687\\
Out[2]=~-------~+~-------~+~-------~+\\
\rule{0mm}{0mm}~~~~~~~~~~~~~3~~~~~~~~~2~~~~~\$eps\\
\rule{0mm}{0mm}~~~~~~~~~\$eps~~~~~~\$eps\\
~\\
\rule{0mm}{0mm}~~~~~~~~~~~~~~~~~~~~~~~~~~~-15\\
>~~~~((22.6599~-~2.36848~10~~~~I)~+~U4sub[1.,~1.,~2.,~2.])~+~O[\$eps]\\[-2ex]
\end{quote}
Here {\tt U4sub} is the function $U_{4,\rm sub}$ introduced in
eq.~\eqref{eq:u4sub}. The list of all symbols returned by {\tt PrepInt} is given
in the last column of Tab.~\ref{tab:fnames}.

{\tt UCall} invokes the numerical code {\tt ucall} (see previous subsection)
to evaluate the finite
remainder functions in the output of {\tt PrepInt}. For the example above this
leads to
\begin{quote}
\tt\small \rule{0mm}{0mm}\\[-2ex]
In[3]:=~UCall[\%]\\
\\
\rule{0mm}{0mm}~~~~~~~~1.66667~~~2.61371~~~7.48687~~~~~~~~~~~~~~~~~~~~~~~~~-15\\
Out[3]=~-------~+~-------~+~-------~+~(-2.36076~-~2.36848~10~~~~I)~+~O[\$eps]\\
\rule{0mm}{0mm}~~~~~~~~~~~~~3~~~~~~~~~2~~~~~\$eps\\
\rule{0mm}{0mm}~~~~~~~~~\$eps~~~~~~\$eps\\[-2ex]
\end{quote}
\pagebreak
Technically, the executable {\tt ucall} is called through an external operating
system command, using the Mathematica function {\tt Run}. The Mathematica
function {\tt UCall} looks for the executable {\tt ucall} in the directory given
by the variable {\tt \$Directory}, which by default is set to {\tt "./"}. For
passing input and output to and from the executable, {\tt UCall} uses the
filenames specified in the variables {\tt \$FileIn} and {\tt \$FileOut},
respectively. In most cases, the user will not need to change any of these
global variables.

Examples for the evaluation of various cases of $U_{4,5,6}$ integrals are
shown in the file {\tt example.m}, which is provided with the TVID program
package.


\subsection{Installation}
\label{sc:inst}

TVID is available for download as a gzipped tar-ball. After saving it in the
desired directory, it can be unpacked by the command
\begin{quote}
\tt tar xzf tvid.tgz
\end{quote}
After unpacking, the numerical C program must be compiled. A make file is
provided for this purpose, which can be invoked with the simple command
\begin{quote}
\tt make
\end{quote}
The make file provided in version 1.0 has been tested on Scientific Linux 6. For
other UNIX-type operating systems, the user may need to change the name of the
C++ compiler (variable {\tt CC}) or explicitly require the inclusion of certain
standard libraries (variable {\tt LIBS}) in the make file. The authors cannot
guarantee that the installation process is successful on any operating system,
but they appreciate any helpful suggestions, comments and bug reports.


\section{Tests}

The output of TVID have been compared to existing results in the literature for
integrals with one and two independent mass scales
\cite{Broadhurst:1998rz,Chetyrkin:1999qi,Grigo:2012ji}, At least ten digits
agreement was found for all tested cases. A similar level of agreement was
reached
for a few comparisons of cases with maximal number of different mass scales with
the work of Ref.~\cite{3vac}.

The typical evaluation times of the numerical module ({\tt ucall}) 
on a single core of an Intel Xeon CPU with 3.7 GHz are: 
$\sim 0.1\,$s for $U_{4,\rm sub}$; $\lesssim 0.01\,$s for $U_{5,\rm sub,0}$;
and 20--30$\,$s for $U_{6,\rm sub,0}$.


\section*{Acknowledgments}

\noindent
The authors gratefully acknowledge communications with E.~de~Doncker, S.~P.~Martin and
D.~G.~Robertson. 
This work has been supported in part by the National Science Foundation under
grant no.\ PHY-1519175.


\end{document}